\documentclass[aps,twocolumn,prl,showpacs,preprintnumbers,
superscriptaddress]{revtex4-1}

\usepackage{amsmath,amssymb,bm,comment}
\usepackage{graphicx}
\usepackage{epsfig}
\usepackage{epstopdf}
\usepackage{hyperref}
\usepackage{longtable}

\begin{document}

\title{Semi-Holographic Quantum Criticality}
\author{Kristan Jensen}
\email{\tt kristanj@uvic.ca}
\affiliation{Department of Physics, University of Victoria,
Victoria, BC V8W 3P6, Canada}
\affiliation{Department of Physics, University of Washington, Seattle,
WA 98195-1560, USA}

\date{\today}
\begin{abstract}
We identify the near-critical effective theory (EFT) for a wide class of low-temperature phase transitions found via holography. The EFT is of the semi-holographic type and describes both holographic Berezinskii-Kosterlitz-Thouless (BKT) and second-order transitions with non-trivial scaling. It is a simple generalization of the Ginzburg-Landau-Wilson paradigm to systems with an emergent (or hidden) conformal sector. Having identified the near-critical EFT, we explore its basic phenomenology by computing critical exponents and low-frequency correlators.
\end{abstract}

\maketitle

\emph{Introduction.}---%
Phase transitions are a central concept in modern physics. The Ginzburg-Landau-Wilson (GLW) paradigm of critical phenomena~\cite{Wilson:1973jj} has proven extremely useful in understanding continuous transitions. However, there are many interesting questions beyond the GLW paradigm~\cite{SVBSF}. For example, a correct model for the low-temperature phases of heavy fermion systems near quantum critical points requires new tools~\cite{2008NatPh...4..186G,2001Natur.413..804S}.

Holography~\cite{Maldacena:1997re,Gubser:1998bc,Witten:1998qj} has become a standard tool to study strong coupling problems. It also offers a road beyond the GLW paradigm. Naively, this should not be so: theories that admit simple dual gravitational descriptions have a large $N$ parameter that suppresses quantum fluctuations in the bulk and field theories. Connected correlators factorize and so the low-energy physics of these theories may be reproduced from a classical EFT. As a result large $N$ theories generally exhibit first-order~\cite{Witten:1998zw} or second-order transitions with mean-field exponents~\cite{Hartnoll:2008vx}.

However, we may turn this observation on its head. Any non-trivial scaling in a holographic phase transition is evidence that the near-critical EFT is outside the GLW paradigm. Indeed, there are already several such examples in the literature. We will focus on two large classes in this Letter. The first are holographic Berezinskii-Kosterlitz-Thouless (BKT) transitions~\cite{Kaplan:2009kr,Jensen:2010ga,Iqbal:2010eh,Jensen:2010vx} and the second are second-order transitions with non-mean-field exponents~\cite{Faulkner:2010gj,Evans:2010np}.

We would like to identify an EFT for these transitions. The way forward was paved in~\cite{Faulkner:2010tq}, which identified the low-energy EFT for a class of holographic non-Fermi liquids. A fermion-fermion two-point function was previously computed in~\cite{Faulkner:2009wj}; the EFT in~\cite{Faulkner:2010tq} reproduces this result ``semi-holographically.'' It is
\begin{equation}
\label{semiPsi}
S_{\rm semi-holo.}=\int dt\,d^dx\left( \mathcal{L}_{\rm free}[\psi]+\psi\Psi_{\mathbf{x}}+\mathcal{L}_{0+1}[\mathbf{x}]\right),
\end{equation} 
where $\psi$ is a free $(d+1)$-d fermion that mixes with a $(0+1)$-dimensional fermion $\Psi_{\mathbf{x}}$ in a large $N$ theory with $(0+1)$-d scale invariance localized at $\mathbf{x}$. Large $N$ guarantees that the two-point function of $\psi$ only depends on the two-point function of $\Psi_{\mathbf{x}}$, which is determined by the $(0+1)$-d scaling. The $(0+1)$-d sector is best thought of as an emergent conformal sector in the IR, realized in the gravitational dual via a near-horizon AdS$_2$ region.

This structure is generic in holography. Semi-holographic EFTs have been derived describing transport~\cite{Nickel:2010pr} and scalars~\cite{Faulkner:2010jy}, mixing light fields with an IR sector. Notably, all of the transitions examined in this Letter rely upon the physics of such an IR theory. For example, holographic BKT transitions\footnote{The name ``holographic BKT transition'' is a misnomer. See~\cite{Jensen:2010vx} for discussion on this point.} are triggered by destabilizing a conformal sector with $(0+1)$-d scaling through the mechanism of~\cite{Kaplan:2009kr}. In the bulk this is accomplished by adjusting control parameters so that the mass of the bulk field dual to the order parameter drops below the Breitenlohner-Freedman bound~\cite{Breitenlohner:1982jf} in the IR region. It is then natural that a semi-holographic theory describes these transitions. We will show that this is the case by matching to previously obtained correlators.

This program is useful for at least three reasons. First, it gives a new class of critical EFTs that may be realized in Nature without any reference to holography. Second, it generalizes the notion of ``universality'' to include the data of an emergent conformal sector. Finally, a number of results that could be computed in holography with great effort may be easily seen from the EFT. \\ \emph{Note:} While finishing this work, we became aware of the related work~\cite{Iqbal:2011aj}. Their results agree with ours.

\emph{Effective field theory.}---%
Our goal is to write down an EFT that reproduces the near-critical one and two-point functions obtained in~\cite{Jensen:2010ga,Iqbal:2010eh,Jensen:2010vx,Faulkner:2010gj,Evans:2010np}\footnote{For recent work toward the same goal, see~\cite{Evans:2011zd}}. The result is
\begin{equation}
\label{semiHolo}
S_{\rm EFT} =\int dt\,d^dx \left(\mathcal{L}_{\rm GL}[\varphi_{i},J_{i}] +  \mathcal{L}_{\rm IR}[\varphi_{i};\mathbf{x}]\right).
\end{equation}
Some remarks are in order. The first term is a standard Ginzburg-Landau action analytic in powers of fields $\varphi_i$ and derivatives with a source $J_i$. For us the $\varphi_i$ are an order parameter, transforming in a representation of a symmetry group. For simplicity, we henceforth suppress the flavor index $i$. The second term is the non-trivial output of holography: $\varphi$ couples to a large $N$ conformal IR sector. As in Eq.~(\ref{semiPsi}), $\varphi$ only couples through mixing: it is a source for a scalar operator in the emergent theory, $\mathcal{O}_{\Delta}$, of dimension $\Delta$. This is a relatively universal feature of holographic systems, whether the IR is a theory with Lifshitz scaling, locally critical scaling, or otherwise~\cite{Faulkner:2010jy}. Finally, Eq.~(\ref{semiHolo}) was previously written down in~\cite{Faulkner:2010jy}, where it was used to great effect. However it was not fully explored: it was thought there to only describe a restricted subset of transitions.

There are three equivalent routes to Eq.~(\ref{semiHolo}). The first is the standard EFT approach, where by trial and error we write down the simplest action that respects the symmetries of the problem while reproducing the correlators of the microscopic theory. The second method follows~\cite{Nickel:2010pr}, where we simplify the dual gravitational problem by decomposing the bulk action into UV and IR sectors that couple to each other other at a radial cutoff $r_{\Lambda}$. We need only know that the UV theory is a gapped, dissipationless theory that, at low energies, is described by the lightest scalar mode, $\varphi$, in the spectrum of the bulk field dual to the order parameter. The UV theory is then a GL model for $\varphi$. Also, $\varphi(r_{\Lambda})$ acts as a source for $\mathcal{O}_{\Delta}$ in the IR theory. Thus we do not need to know the details of $\mathcal{L}_{\rm IR}$ to compute correlators of $\varphi$, just its response. Lastly, we may integrate out bulk fluctuations down to the IR of the dual geometry to obtain an EFT in terms of a scalar operator in the IR sector with various multi-trace deformations~\cite{Faulkner:2010jy}. In terms of Eq.~(\ref{semiHolo}) we may integrate out $\varphi$ to write an EFT in terms of $J$ and the operators of $\mathcal{L}_{\rm IR}$. For example, the quadratic coupling of the GL term maps to a double-trace deformation for $\mathcal{O}_{\Delta}$.

The last approach is particularly useful as it naturally addresses the question of quantization in the IR sector. For generic $\Delta$, the IR has one fixed point. However for $\Delta \in \left(\frac{1}{2},\frac{3}{2}\right)$ the conformal theory has an alternate quantization in which $\mathcal{O}_{\Delta'}$ has dimension $\Delta'= 1-\Delta$. This is a little unnatural when writing an effective theory for $\varphi$, but if instead we integrate out $\varphi$ we find an effective IR theory with a relevant double-trace in which the critical physics is more manifest. We will return to this shortly.

\emph{Second-order transitions.} We will now compute correlators of the EFT Eq.~(\ref{semiHolo}) for second-order transitions where the IR sector has $0+1$-dimensional scaling symmetry. The IR sector is then ``locally quantum critical''~\cite{2001Natur.413..804S} and is encoded in the gravity dual by a near-horizon AdS$_2$ region. However, this is not the only instance where our EFT correctly reproduces holographic results -- it is simply an instructive example. Moreover, we will temporarily take $\Delta >1/2$, the standard quantization. For definiteness, we expand the GL action as
\begin{equation}
\label{GLS}
\mathcal{L}_{\rm GL}=-\frac{c_t^2}{2}(\partial_t\varphi)^2+\frac{c_x^2}{2}(\partial_x\varphi)^2+\frac{c_2}{2}\varphi^2+\frac{c_4}{4}\varphi^4+\hdots
\end{equation}
where the couplings are analytic functions of external control parameters. We could study a theory with a charged order parameter at nonzero chemical potential by making the derivatives gauge-covariant derivatives. As in the usual GL framework, a second-order transition is triggered when $c_2$ is tuned to vanish, provided that $c_4>0$. We parametrize $c_2$ near the transition as $g-g_c$, for $g$ some control parameter. In the framework of~\cite{Faulkner:2010jy} where $\varphi$ is integrated out, this corresponds to the double-trace deformation of the IR theory changing sign from positive to negative. Let us now compute one-point functions in the broken phase. Assuming a translationally-invariant equilibrium state, we simply solve $\delta \mathcal{L}/\delta \varphi=0$,
\begin{equation}
\label{extremize}
(g-g_c)\langle \varphi\rangle + c_4\langle \varphi^3\rangle + \hdots+\langle \mathcal{O}_{\Delta}\rangle =0.
\end{equation}
Now, recall that $\varphi$ acts as a source for $\mathcal{O}_{\Delta}$ and so we should assign $\varphi$ dimension $1-\Delta$  in the IR theory. In this counting the mixing term is marginal. At zero temperature, scale invariance in the IR theory then implies that $\langle \mathcal{O}_{\Delta}\rangle\propto \langle\varphi\rangle^{\frac{\Delta}{1-\Delta}}$. As a result the condensate $\langle \varphi\rangle$ scales differently in the broken phase $g<g_c$ depending on the value of $\Delta$. For $\Delta\geq \frac{3}{4}$, the quartic term is relevant and so dominates the mixing term so that $\langle\varphi\rangle \propto (g_c-g)^{\beta}$ with $\beta=1/2$. However, for $\Delta\in \left( \frac{1}{2},\frac{3}{4}\right)$, the quartic term is irrelevant and the exponent $\beta$ becomes $\frac{1-\Delta}{2\Delta-1}$.

At nonzero temperature, $\langle \mathcal{O}_{\Delta}\rangle$ scales differently with the source $\varphi$. We now have $\langle \mathcal{O}_{\Delta}\rangle = T^{\Delta}v\left(\langle\varphi\rangle /T^{1-\Delta}\right)$, where $v$ is an odd analytic function. Then $\beta$ takes the mean-field value $1/2$ for all $\Delta$. However, the transition is displaced by $g_c(T)=g_c-v'(0)T^{2\Delta-1}$. The non-mean-field scaling at $T=0$ then persists as a $T$-dependent scaling in the location of the transition. Alternatively, $T_c$ scales as $|g_c(T)-g_c|^{1/(2\Delta-1)}$.

The dynamical susceptibility of $\varphi$, $\chi(\omega,\mathbf{k})$, may be computed by factorization to be
\begin{equation}
\label{2ptFactor}
\chi(\omega,\mathbf{k})= \frac{1}{\chi_0(\omega,\mathbf{k})^{-1}-\mathcal{G}_0(\omega,\mathbf{k})},
\end{equation}
where $\chi_0$ and $\mathcal{G}_0$ are the two-point functions of $\varphi$ and $\mathcal{O}$ at zero mixing. In the normal phase, $\mathcal{G}_0$ takes different forms at $T=0$ and $T>0$,
\begin{equation}
\label{2ptIR}
\mathcal{G}_0=\left\{ \begin{array}{ll} f_0(-i\omega)^{2\Delta-1},\hspace{.3cm} & T=0, \\ T^{2\Delta-1}f\left(\frac{ \omega}{T}\right),\hspace{.3cm} & T>0, \end{array}\right.
\end{equation}
where $f(x)$ is analytic. Then $\chi$ is approximately
\begin{equation}
\chi\approx \frac{1}{g-g_c + c_s^2 k^2-c_t^2\omega^2-f_0(-i\omega)^{2\Delta-1}}
\end{equation}
at $T=0$, which has a low-frequency pole that lives on the imaginary axis at zero momentum. This mode is in the lower-half-plane (LHP) for $g>g_c$ and wanders into the upper half plane (UHP) for $g<g_c$, driving the transition. We then find that the correlation length diverges as $\xi\sim (g-g_c)^{-\nu}$ with $\nu=1/2$. Similarly, the correlation time diverges as $(g-g_c)^{-z\nu}$ where $z$ is
\begin{equation}
z=\left\{ \begin{array}{ll} \text{Max}\left[ 1,\frac{2}{2\Delta-1}\right], \hspace{.3cm} & T=0, \\
2, \hspace{.3cm} & T>0.\end{array} \right.
\end{equation}
Also, $f(0)$ will in general be nonzero for $T>0$. This shifts the transition by $g_c(T)=g_c+f(0)T^{2\Delta-1}$, matching the $T$-dependent scaling found above.
\begin{longtable*}{c|cccccccc}
 & $\alpha$ & $\beta$ & $\gamma$ & $\delta$ & $\nu$ & $z$ & $\eta$ & $T_c$ \\ \hline
$T=0$: & $\text{Min}\left[ 0,\frac{4\Delta-3}{2\Delta-1} \right]$ & $\text{Max}\left[\frac{1}{2},\frac{1-\Delta}{2\Delta-1}\right]$ & $1$ &$\text{Min}\left[ 3,\frac{\Delta}{1-\Delta} \right]$ & $\frac{1}{2}$  & $ \text{Max}\left[1,\frac{2}{2\Delta-1}\right]$ & $0$ \\ $T>0$: & $0$ & $\frac{1}{2}$ & $1$ & $3$ & $\frac{1}{2}$ & $2$ & $0$ & $\frac{1}{2\Delta-1}$
\\ 
\caption{\label{critExpC1} The exponents of the semi-holographic EFT when the IR sector is locally quantum critical.}
\end{longtable*}
By similar methods we also compute the remaining critical exponents, $\alpha, \gamma, \delta,$ and $\eta$. Our results are summarized for the locally critical case in Table~\ref{critExpC1}. We have also computed the case where the IR sector has $(d+1)$-dimensional scaling symmetry. In each case the exponents satisfy the scaling relations~\cite{Itzykson:1989sx}
\begin{equation}
\gamma=\beta(\delta-1) = \nu(2-\eta)=2\beta\delta+\alpha-2,
\end{equation}
but not so-called ``hyperscaling'', $\alpha = 2-\nu d$. This result should not surprise us: the exponents of classical models do not depend on dimension and so do not hyperscale.

We may also compute the two-point function of $\varphi$ in the ordered phase by factorization, giving Eq.~(\ref{2ptFactor}). However, we now interpret $G_0$ as the two-point function of $\varphi$ in $\mathcal{L}_{\rm GL}$ with a source $g\langle \mathcal{O}_{\Delta}\rangle$ and $\mathcal{G}_0$ as the two-point function of $\mathcal{O}_{\Delta}$ in $\mathcal{L}_{\rm IR}$ in the presence of a source $g\langle \varphi\rangle$. It follows that  $G_0^{-1}\sim c_2'+c_x^2k^2-c_t^2\omega^2+\hdots$ for some $c_2'\neq c_2$ a function of control parameters.

We return to the question of the near-critical physics in the alternate quantization, $\Delta' \in (-\frac{1}{2},\frac{1}{2})$. The dynamical susceptibility at zero temperature is now
\begin{equation}
\label{altChi}
\chi\approx \frac{Z+Z_0 \mathcal{G}_0(\omega)+\hdots}{(g-g_c+\hdots)\mathcal{G}_0(\omega)-a+\hdots},
\end{equation}
which is also the form of the two-point function of $\mathcal{O}_{\Delta'}$ in the presence of a double-trace deformation $g-g_c$. Note that there is no light singularity near the transition $g=g_c$. Rather $g<g_c$ switches the sign of the spectral function, triggering a quantum phase transition. We should expect this result: in the effective IR theory, $g>g_c$ corresponds to a positive double-trace deformation and $g<g_c$ to a negative one, which destabilizes the theory to a condensed phase.

\emph{Special cases.}---The EFT Eq.~(\ref{semiHolo}) admits richer phenomenology than discussed above. For now we focus on the extreme case $\Delta=1/2$, where a number of exponents diverge. As discussed below, the system is on the precipice of a holographic BKT transition, and so we may regard it as a multicritical point at which both $c_2$ and $\Delta$ are tuned to critical values. Notably the standard and alternate quantizations coincide, so a double trace is marginal at $c_2=0$. Following~\cite{Iqbal:2011aj} we term this multicritical point a ``Marginal quantum critical point (QCP).'' However, taking $g<g_c$ condenses the GL sector, which will condense $\langle\mathcal{O}_{\Delta}\rangle$ so that a negative double trace ($g<g_c$) is marginally relevant. Now consider the dynamic susceptibility. Since the negative double trace is marginally relevant, the two-point function is of the form Eq.~\ref{altChi}.  At $T=0$ the Green's function of the IR theory is $\mathcal{G}_0\sim \ln (-i\omega)$, while for $T>0$ it is $\mathcal{G}_0\sim f(\omega/T)$, giving
\begin{equation}
\label{marginalChi}
\chi\approx \frac{Z_1 \ln \left( -\frac{i\omega}{\omega_1}\right)+\hdots}{(g-g_c) \ln \left( -\frac{i\omega}{\omega_0}\right)-1},
\end{equation}
at $T=0$. In the broken phase $g<g_c$ there is an unstable mode with $\omega = i\omega_0 \exp\left(\frac{1}{g-g_c}\right)$. This mode approaches $\omega = 0$ exponentially fast as $g\rightarrow g_c$, corresponding to the generation of an exponential scale $\Lambda \sim \exp\left(\frac{1}{g-g_c}\right)$. We should then expect an exponentially suppressed condensate $\langle \varphi\rangle\sim\Lambda^{1/2}$, which should scale as $\Lambda^{1/2}$ as $\varphi$ is dimension $1/2$ in the IR theory. It is interesting that the form of the bosonic susceptibility Eq.~(\ref{marginalChi}) is the same as the fermionic susceptibility hypothesized to describe marginal Fermi liquid (MFL) phases~\cite{Varma:1989zz}. Curiously, this form may also be obtained for both bosonic and fermionic fluctuations in a recent holographic model for a MFL~\cite{Jensen:2011su}. Indeed, that model also offers a string theory embedding of a marginal QCP.

\emph{Holographic BKT.}---%
There is another way to trigger a transition in this theory. We may break the scale invariance of the IR theory by driving $\Delta$ into the complex plane at $T=0$, which condenses $\langle\mathcal{O}_{\Delta}\rangle\neq 0$. This vev acts as a source for $\varphi$ and so condenses $\langle\varphi\rangle$ as well. In renormalization group (RG) language, this may be done by the collision and annihilation of the standard and alternate quantizations, which may then bifurcate into the complex plate~\cite{Kaplan:2009kr}. This corresponds to an operator dimension that depends on control parameters $g$ as $\Delta_{\pm} = \frac{1\pm\sqrt{g-g_c+\hdots}}{2}$. The square root is important -- $(\Delta - 1/2)^2$ must be analytic in $g$ in order to have two fixed points. Indeed this structure has been found in a number of models~\cite{Jensen:2010ga,Kutasov:2011fr}. Conformality is lost for $g<g_c$ as the scale invariance of $\mathcal{L}_{\rm IR}$ is broken. The RG analysis of~\cite{Kaplan:2009kr} showed that for small $g_c-g\ll 1$ there is a spontaneously generated scale $\Lambda\sim \Lambda_{\rm UV}\exp\left(-2\pi/\sqrt{g_c-g}\right)$ with $\langle \mathcal{O}_{\Delta}\rangle\sim \Lambda^{\frac{1}{2}}$, for $\Lambda_{\rm UV}$ a UV scale. Eq.~(\ref{extremize}) then gives $\langle \varphi\rangle \sim \exp {-\pi / \sqrt{g_c-g}}$. Actually, there is an infinite tower of scales $\Lambda_n\sim \Lambda_{\rm UV}\exp (-2\pi n/\sqrt{g_c-g})$ generated, giving states with $\langle \mathcal{O}_{\Delta}\rangle_n \sim \Lambda_n^{\frac{1}{2}}$ and so $\langle\varphi\rangle_n\sim \exp(-n\pi/\sqrt{g_c-g})$. These are the `Efimov extrema' of~\cite{Jensen:2010ga}.

The two-point function of $\varphi$ is just Eq.~(\ref{2ptFactor}) for the normal phase. There is one important thing to note: this transition is not triggered by tuning $c_2$, but rather the dimension of $\mathcal{O}_{\Delta}$. In fact, $c_2$ must be positive in order to be in the normal phase near $\Delta = 1/2$. We now consider this two-point function in two limits. First, at exactly $\omega=0$ the numerator and denominator of Eq.~(\ref{2ptFactor}) are analytic in $2\Delta-1$, which goes as $\sqrt{g-g_c}$ near the transition. Thus the static susceptibility goes as
\begin{equation}
\chi(\omega = 0,\mathbf{k})\approx \frac{b_0+b_1\sqrt{g-g_c}}{a_0+a_1 \sqrt{g-g_c}},
\end{equation}
which is (i.) finite at the transition, in contrast with transitions in the GLW framework, and (ii.) has a square root singularity at $g=g_c$. Second, we look at $\chi$ in the small frequency limit. One can easily show that, near $g=g_c$ in the normal phase, $\chi$ is
\begin{equation}
\chi(\omega,\mathbf{k})\approx \frac{Z\sinh \left( \sqrt{g-g_c}\ln\left(-\frac{i\omega}{\omega_1}\right) \right)}{\sinh \left( \sqrt{g-g_c}\ln\left(-\frac{i\omega}{\omega_0}\right) \right)},
\end{equation}
with $\omega_0>0$. In the normal phase $\chi$ has no light poles, while in the broken phase there is an infinite tower of unstable modes with $\omega_n = i\omega_0 \exp( -2\pi n/\sqrt{g_c-g})$. One may also study the $T>0$ physics of the transition by using the $T>0$ result for $\mathcal{G}_0$, Eq.~(\ref{2ptIR}) with Eq.~(\ref{2ptFactor}).

\emph{Relation to previous work.}---%
The exponential scaling of $\langle\varphi\rangle$ in the holographic BKT transitions observed here matches that seen in~\cite{Jensen:2010ga,Iqbal:2010eh,Jensen:2010vx}. The two-point function of $\varphi$ in the disordered phase also matches results in the literature~\cite{Faulkner:2010gj,Evans:2010np}. The behavior of the correlation length and static suceptibility, however, are predictions. Finally, the near-critical scaling of the holographic BKT transition identified in~\cite{Edalati:2011yv} may be captured by our EFT when the IR theory has anisotropic (Lifshitz) scaling. 

For second-order transitions, the behavior of the one and two-point functions of the condensate $\varphi$ computed above for the locally critical case matches those found in holographic superfluids at $\mu\neq 0$~\cite{Faulkner:2010gj} as well as probe brane systems~\cite{Evans:2010np}. The case where the IR sector has $(d+1)$-d scale invariance may also describe holographic superfluids at $\mu=0$ with a negative double-trace~\cite{Faulkner:2010gj}.

\emph{Discussion.}---%
We have seen that the near-critical physics of a wide class of holographic phase transitions may be understood with a classical EFT. This is non-trivial in itself. Now we make a few observations on what we have learned. First, the second-order transitions are a mild deviation from the usual GLW paradigm. The non-trivial critical exponents are essentially a consequence of nonzero anomalous dimensions in the IR sector. The marginal and holographic BKT transitions are a bit more interesting. They exhibit the generation of exponentially suppressed scales, but in each the normal phase is gapped and the  near-critical static susceptibility is finite. This result is generic; it does not rely upon local criticality in the IR sector. Instead of the growth of long-distance fluctuations of symmetry-breaking operators, these transitions are triggered by the breaking of conformal symmetry in the emergent sector, which then condenses the emergent fields and so the order parameter. This is a qualitatively new mechanism for criticality.

There is one issue we have ignored thus far. Most holographic systems with a near-horizon AdS$_2$ factor exhibit a variety of low-temperature instabilities (e.g.~\cite{Hartnoll:2008vx}), so that the physics of a locally critical sector only exist over a range of energy scales. Indeed, there is a recent argument that locally critical sectors are \emph{always} unstable at sufficiently low energies~\cite{Jensen:2011su}. However, all is not lost. One can take the outlook of~\cite{Iqbal:2011in}, in which our Eq.~(\ref{semiHolo}) is a good description over an intermediate range of energies and temperatures. Alternatively, Eq.~(\ref{semiHolo}) may be used as a starting point for describing the \emph{lowest} energy scales, provided that the IR sector is not locally critical but rather some other interesting scale-invariant theory.

Many interesting questions remain. Most of the interesting results in the GLW paradigm arise from quantum corrections -- what is the quantum phenomenology of Eq.~(\ref{semiHolo})? Can the $1/N$ effects of the bulk theory be computed by including the $1/N$ physics of the IR sector? 

\emph{Acknowledgments.}---It is a pleasure to thank Tom Faulkner, Carlos Hoyos, Andreas Karch, Hong Liu, and Dam Son for many useful discussions. This work was supported in part by the U.S. Department of Energy 
under Grant Numbers DE-FG02-96ER40956 and DE-FG02-00ER41132 as well as NSERC, Canada.

\bibliography{refs_v2}
\end{document}